# Aggregated Massive Modular Paradigm: A 6G Telecom Infrastructure Vision

Laurent PETIT, *Senior Member, IEEE*, Jean-Baptiste DORE, *Member, IEEE,*
Eric MERCIER, Claude BROCHETON, Julien LEGRAND and Dimitri KTENAS, *Member IEEE*

*Abstract*— **On the eve of 5G network deployment worldwide, a new boom in the mass digitization of our personal and professional lifestyles is looming. 5G marks a strong shift in the way in which connectivity will impact our societies and industries. This shift is taking place through a combination of technologies that have progressed business-as-usual with technical improvements. The future of 5G, already called 6G, currently belongs to the world of research, which must imagine the future development of mobile digital services in 10 years. In this ever more connected world, ownership of data and protection, control of ecological impact, both energetic and electromagnetic, in addition to guarantee an available, flexible network will be essential factors in the success of 6G. This new generation of networks brings along great challenges in terms of modularity, cost effectiveness and power efficiency. We present here a novel disruptive deployment strategy using cost-efficient modules along with aggregated connectivity meeting these requirements.**

## I. INTRODUCTION

THIS document is an open concept presentation of a Massive Modular Paradigm (MMP) paving the way for future pervasive network towards fully virtualized, dedicated to the next generation of telecommunication infrastructures. As leader in the design, development and manufacturing of innovative interconnection solutions, Radiall is an enabler for connectivity technologies and forward-compatible solutions for new architecture. From our point of view, subsystems for 6G platforms would require disruptive innovation in interconnect and microwave-to-millimeter wave (mmW) component solutions up to the THz spectrum in order to reach the new subsystems architecture integration challenge. Radiall and CEA-Leti have created a common laboratory on RF and Photonic topics, targeting '5G & 6G' development [1], working on Hardware 'disruptive' technologies for 5G and beyond. On the telecommunication market, Radiall is developing high-end cost-efficient solutions based on coaxial and optical connectivity for outside and inside Remote Radio Head (RRH). Radiall teams have extended out horizon working on telecom infrastructures, mmW antennas for X-Haul and access, tunable and reconfigurable filters and antennas.

As there are some better insights on radio technology system definition and identified technology components for an overall 6G-system architecture, we would like to share our 6G vision compatible with ubiquitous modularity for future mobile networks. The vision depicted in this paper is based on cost efficient modules along with aggregated connectivity. The paper is organized as follow. In section II, a vision for 6G is proposed and discussed. Section III gives a description of key technologies and more precisely how cost efficient modules will be at the heart of 6G. Section IV highlights the aggregated connectivity framework. Last, section V concludes this paper.

## II. PROPOSED 6G VISION

Many efforts have been deployed to change the cellular industry paradigm to design a new 5G network enabling new services that go beyond the generalization of access to broadband internet access everywhere. 5G New Radio (NR) paves the way of new applications with heterogeneous requirements, throughput, latency, connection density, reliability or lifetime… and with the ambition to target vertical markets. In addition, new frequency bands are being considered in order to support the traffic growth, with the major breakthrough which will consist in network virtualization. Many experts mention that the next big step for cellular networks lies in their cloudification that will support the explosion of radically new services, rather than 5G.

In our shared vision, the next generation of wireless systems will transform the 5G service-oriented networks into user-centric and machine ad-hoc dynamic (re)configuration of network slices. This will be made possible by software-defined end-to-end solutions from the core to the radio access network (RAN), including the air interface [2] as well as the RF and antenna systems [3]. These latter are envisioned as key technologies to meet the user/service requirements.

The demand for reliable wireless connectivity providing 100% availability will be a holy grail for the industry and the attractiveness of a given geography. Despite the effort made by the 5G standardization body, we believe that the basic concept of cellular connectivity should be revisited to satisfy this reliability constraint. On the end-point mobile side, the

---

multi-connectivity scheme (frequency, time and space) is a way to increase the reliability of a wireless transmission with still many challenges to solve. From the infrastructure side, the migration to a mesh network system ensures the reliability of the infrastructure. The first consequence is the need of extreme xHaul capabilities (bandwidth, latency requirement). This requirement makes fiber solutions desirable, but sometimes complicated by local installation constraints to efficiently deploy networks and enable penetration. The wireless infrastructure is foreseen as a complement to the optical fiber deployment as it offers more agility, shorter installation times, and in the case of a mesh architecture strong reliability. It may also provide connectivity to mobile, removable or even flying access points. Last but not least, these solutions will be valuable if and only if their cost efficiency is demonstrated.

From a technological perspective, this paradigm will require wireless mmW links in V-, E-, W-, D-bands namely 60 GHz to 81 GHz, 90 GHz and 140 GHz. Moreover, configurable antenna enabling alignment capabilities, will be needed and will be empowered by antenna-RF co-integration as well the introduction of digital and software processing as close as possible to the radiating element [3]. At the edge of the network, the design of cost- and power-efficient RRH is of particular interest. 5G has introduced new paradigm with the use of large collocated antenna panels (typically 64 TRX), which are currently deployed and still facing some challenges on optimization and reliability. Indeed, the massive MIMO concept proves itself to be both complex and expensive while being particularly power-intensive: such a sub-optimized hybrid approach would be intolerable from the perspective of both cost- and power-efficiency. These elements should not to be underestimated on a competitive market and growing 'greener' consciousness. Besides mobile subscribers would accept to increase their fees paid to operators to absorb huge 'Massive MIMO' capital expenditure (CAPEX).

The rise in data throughput is greatly alleviated by using many antennas and/or many transmitters or local small-cells [4]. This is clearly a data capacity booster. The first approach prevails for small-cell multiplication while the second approach militates for Massive MIMO architecture. The price to pay in terms of complexity, data processing or power consumption can be variable and sometimes too high. Small-cell multiplication and Massive MIMO are often presented as competitive approach for future infrastructures deployment [5].

OEMs or operators may seek alternative or complementary approach to the Massive MIMO paradigm. The latter can prove itself to be suboptimal in terms of coverage and energy efficiency and often comes at the cost of increased complexity of electronic systems. It would be better to have a modular approach and aggregate 6G modules in an appropriate and optimized manner; according to use cases (data throughput vs number of users) different scenario or geographic situations (urban canyon, dense area, open fields), rather than 'collocating everything' in one specific place.

The next generation of wireless system could move to large collocated antenna array through network virtualization also referred to Distributed, Cooperative and Cell Free MIMO [5] anticipating highly modular for full Virtual Radio Access Network (VRAN). This paradigm fits well with the proposed vision for the next generation of mobile network: the design of a dense network to support the massive deployment of cost and energy-efficient RRH virtually building large antenna array. Despite the increase of coverage, this paradigm could decrease the electromagnetic field (EMF) and allow global optimization of power consumption. On the one hand, Figure 1 depicts the traditional architecture of communication networks based on a cells distribution principle. The main issue of such architectures is the limited Quality-of-Service (QoS) at the cell edge as well as the need of a large antenna panel located at the center of cell. On the other hand, Figure 2 illustrates the envisioned 6G architecture with (i) a micro-cell with few antennas virtually creating a large antenna and (ii) high bandwidth xHaul connections to ensure seamless connectivity between micro-cells. The day-to-day efforts in the development of the Open RAN (Open Radio Access Network) [6] defining standardized interfaces and split, will make distributed and software defined MIMO processing powerful. Moreover such architecture would give the operator a wide range of possibilities in the choice of equipment and would enable the emergence of a multivendor solutions ecosystem compatible with various requirements.

To summarize, the vision, it is envisaged extreme densification of the network infrastructure which will cooperate to form a cell-free network with seamless QoS over the coverage area fed by a high-capacity mesh xHaul link, based on fiber and mmW connectivity with software reconfiguration ability (including dynamic beam-steering).

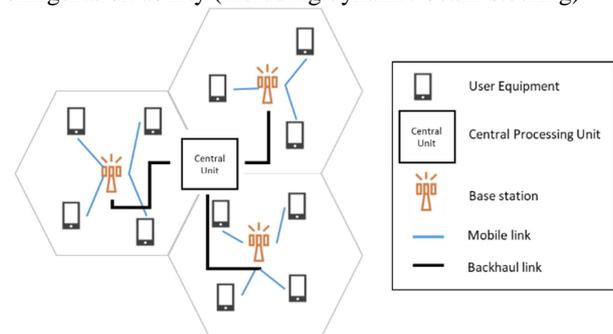

**Fig. 1. 5G paradigm with large panel antenna and cell**

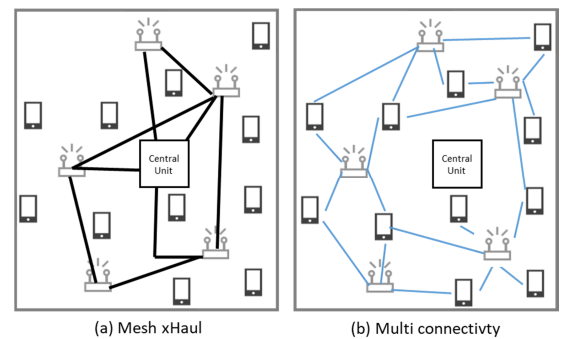

**Fig. 2. Beyond 5G paradigm with multi-connectivity, mesh architecture and a large number of a "few antennas" micro-cell.**

## III. COST EFFICIENT MODULES AS ENABLERS

Our 6G vision anticipates high modularity for full VRAN. It would use many cost-efficient transceiver modules. These 6G modules would have their operating power consumption downed to the watt-level or less, opening up many possibilities as for manufacturing techniques of transceiver components and integrated systems. High power filters conventionally used in 4G/5G base station use bulky, heavy cavity filter technologies; however downscaling power per RF line let us envision other techniques such as ceramic filter or even SAW/BAW filters. Many are commercial off-the-shelf (COTS) or dedicated microelectronics IPs that are already massively used in current mobile User Equipment (UE) such as smartphone produced today by billions of units. These building blocks, namely cost efficient TRX modules, could consequently be derived from current technologies used for sub-6GHz spectrum and be deployed on the network side as depicted in Fig 3.

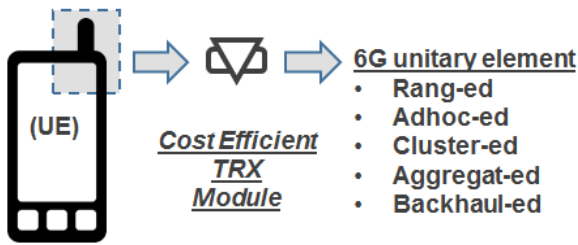

**Fig. 3.** Cost Efficient modules could re-use or derived from mass produced smartphones components

From our perspective, the 6G key hardware enabling technologies are:

- Air interface towards mmW and up to Terahertz
- Full-Duplex HW/SW implementation
- Antenna & filters flexibility / reconfiguration
- Antenna RF microwave and optical convergence

The target in order to reach ultimate VRAN would then rely on reconfigurability at different levels on a broad range of frequencies towards mmW and THz. Full-Duplex HW implementation into the module would double data throughput capacity; and recent research showed that integrating solid state circulator [7] could be a solid option as opposed to classic bulky heavy magnetic architecture.

Antenna beam-switching, -steering or -forming but also filter tunability will be decisive for beyond 5G waveforms and architectures. In order to reach 6G objectives [4] [8] [9] the cost-efficient modules could be "ranged". One way forward could be to scale power (10mW to 10s of W) or use directive antenna-array for specific use case scenario (e.g. urban canyon, streets or indoor corridor). Analog phase-shifting should be considered too. Compact space MIMO diversity antenna using switched parasitic concept such as presented in [10] could be envisaged to enhance spatial diversity and be a great benefit for VRAN deployment. All this tunability could be implemented in standard technologies or more advanced ones like RF MEMS or Phase-Change Materials (PCM) [11], which now show acceptable packaging and industrial yield after three decades of R&D.

Finally one should consider microwave and optical convergence as an opportunity to bring fiber-to-the 6G module using RF-over-Fiber or more classic data over fiber to link modules at optical frequencies with different support (glass or plastic or dielectric waveguide) with much lower loss than electrical coaxial and consequently greater range to haul data back and forth.

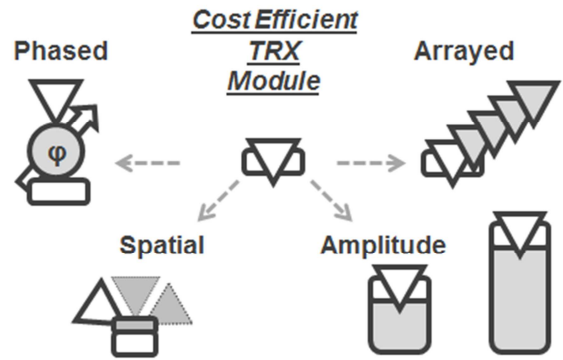

**Fig. 4.** Ranging Cost efficient Module for 6G

## IV. AGGREGATED CONNECTIVITY TOWARDS 6G

As discussed previously, our vision would use many cost-efficient TRX modules to set a highly modular dense mesh network bringing pervasiveness and reconfigurability for VRAN. All these 6G building blocks could share data using a cluster approach in order to collaborate in a VRAN paradigm, i.e. cell-free MIMO approach. Combining AdHoc mode should be considered via a 'master relay' to reach transport/backhaul layers.

In most cases, however, our approach would require an **aggregation connector**, which would locally gather data from a cluster in any topology (inline or stars). This new wired connectivity approach is a clear requirement in order to address the aforementioned challenges and paves the way for ubiquitous QoS of high mobility devices with energy and cost efficiency. This mesh architecture deployment would be enabled by these new connecting interfaces such as an **O**ptical and/or RF **A**ggregation **C**onnector for **L**eading **E**dge Telecom infrastructures (ORACLE). This latter would cover low frequency to optical along with RF microwave and mmW frequencies.

The physical links from aggregation connector to module or networks could use different media from copper to optical through mmW or even using plastic waveguide [12]. The 'Hybrid' approach in-between fully-analog and fully-digital could and should be chosen thoughtfully.

Another aspect would be the functionalization of connectors (SerDes, filters, metering, pre-processing…) which could bring pervasive and distributed computing/controlling to the 6G network; and updated though artificial intelligence.

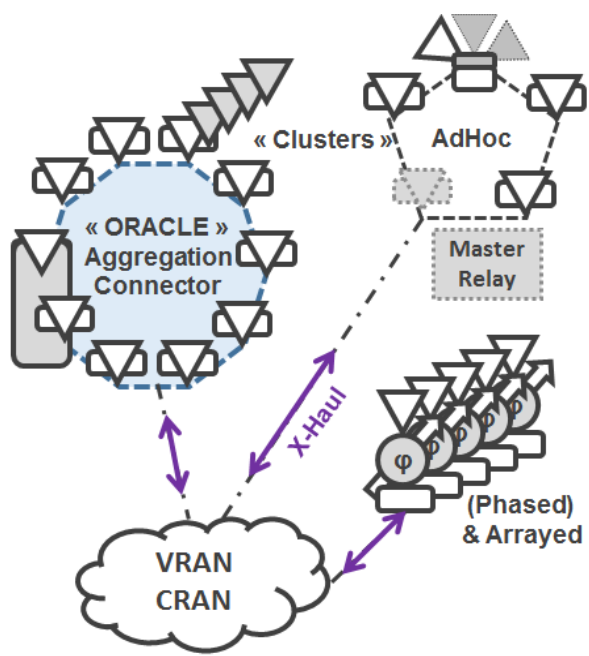

**Fig. 5. ORACLE connector as enabler aggregating 6G Module Aggregation for ubiquitous VRAN deployment**

RF-over-Fiber should be considered for transport, with different variant from DWDM (Dense Wavelength Division Multiplexing), analog Radio-over-Fiber (ARoF), eCPRI (enhanced Common Public Radio Interface) which results from a high-resolution digitization of the radio signals notably involving high bit-rates and time-sensitive signals.

Moreover, the wireless infrastructure is seen as a promising complement to the optical fiber deployment as it offers more agility and shorter installation times. This aggregation point, namely a Point-of-Connection (PoC) would then need to "haul" a certain amount of data (depending of scattering and segregation). A wireless solution could be to use a pencil beam mmW antenna such as Radiall V-band or E-band backhauling antenna solution [13] or even near-THz band [14] to address the integrated access and backhauling (IAB). Additionally it would be useful to implement our auto-pointing innovating mechanism [15] both for low speed install procedure and fast response for mast swing and vibration compensation in windy conditions.

Hardware Key Enabling Technologies will be tunable, ranged and collaborative cost-efficient modules for a 6G VRAN. We believe the "aggregated connectivity", based on wired and wireless links on a broad range of frequencies and topologies, will be the backbone of a successful 6G deployment.

## V. CONCLUSION

Our novel disruptive deployment strategy focuses on using cost-efficient modules along with aggregated connectivity that target 6G data throughput and capacity requirement. From our perspective it would be a great benefit to operators and users. Cost effectiveness will be met through integration and volume. Indeed, cost efficiency would be derived from mass-market of sub-6GHz devices (already mainly COTS) and later toward mmW such as 5G-FR2 and near-THz. Our shared vision provides a great perspective for ubiquitous modularity and pervasive networks which are clear and challenging objectives of 6G roadmaps.

Authors propose a vision and are grateful for open discussions.

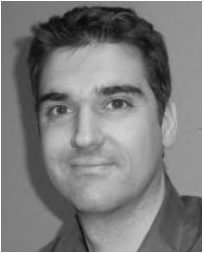

**Laurent PETIT** (S'03–M'06- SM'20) received the M.S. degree in physics and microelectronics engineering from "Télécom Physique Strasbourg", in 2003 and "Microelectronics and Microwave" PhD from Université Grenoble Alpes in 2006. Between 2003 and 2006, he was involved at CEA-LETI, in advanced antenna development of "spatial MIMO"; MEMS based reconfigurable antennas and authored or co-authored several journal and conference papers on his topic. In 2006 he joined CSEM Centre Suisse d'Electronique et de Microtechnique working in Wireless Communication RF sensors; Wireless "Pre-5G" concepts; Microwave sub-6G and Millimeter Wave components and systems. Dr. PETIT joined Radiall in 2011 as Innovative Antenna R&D Engineer. In 2012, he joined Research and Technology group as project manager and Corporate Microwave Expert working on innovative connectors; MMW antennas and Multiphysics with several publications and patents.

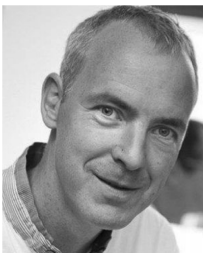

**Jean Baptiste DORÉ** received his MS degree in 2004 from the Institut National des Sciences Appliques (INSA) Rennes, France and his Ph.D. in 2007. He joined NXP semiconductors as a signal processing architect. Since 2009 he has been with CEA-Leti in Grenoble, France as a research engineer and program manager. His main research topics are signal processing (waveform optimization and channel coding), hardware architecture optimizations (FPGA, ASIC), PHY and MAC layers for wireless networks. Jean-Baptiste Doré has published 50+ papers in international conference proceedings and book chapters, received 2 best papers award (ICC2017, WPNC2018). He has also been involved in standardization group (IEEE1900.7) and is the main inventor of more than 30 patents.

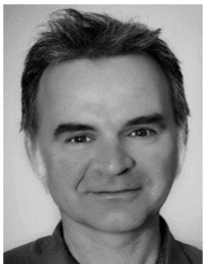

**Eric MERCIER** graduated from the ENSEEIHT of Toulouse, France, in 1991, and holds a DEA in microwaves focused on near-field/far-field antenna diagram conversion done at Alcatel Space, Toulouse, France. After having held positions in the optical test equipment with Schlumberger, for physical fiber optical link tests, as analog and signal processing engineer, he has pursued his work in semiconductor companies like ST and Atmel as R&D Application & Characterization engineer, as well as Marketing engineer. His main field of interest has been low power RF dedicated to Wireless Sensor Network. He is now at CEA-Leti, France, since 2006, where he has led projects in the scope of ULP RF, with a specific focus on low-power RF transceiver design & implementation as well as on embedded resources dedicated to low-power WSN solutions. He is now the Head Manager of the Laboratory for Architectures & Integrated RF design (LAIR). This lab is in charge of designing RF solutions for ULP, UWB, UNB, mmW, high-data rate, RFID, PA, and FEM, with a common target of addressing the lowest possible power consumption and make use of the most advanced CMOS technologies. He has coauthored some conference papers and participated to a book chapter on the topic of WSN.

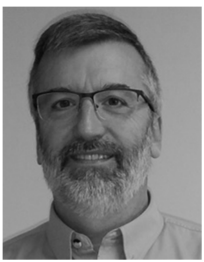

**Claude BROCHETON** received his master degree in mechanics and energetics from the National Institute of Applied Sciences from Toulouse in 1986. He joined the research department of Radiall France in 1988, where he participated to the definition and development of the RF shielding effectiveness (SE) measurement methods by the reverberant chamber and the injection lines. He joined the coaxial connectors R&D department in 1991, and developed the first Passive Intermodulation test measurement equipment in Radiall; He led the Antenna R&D office, first in France, and later in US, from 1994 to 2003 and is currently in charge of the RF Interconnexion R&D offices in France, North-America and Asia since 2007.

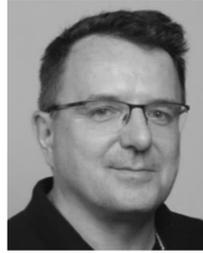

**Julien LEGRAND** received a Master's degree in Fundamental Physics from University of Paris XI Orsay - Saclay in 1995 and his Master of Science degree in Condensed Matter and Soft Matter Physics from the University of Paris VI Jussieu - Sorbonne in 1996 and PhD in the field of magnetic nanomaterials (LM2N Laboratory and CEA Saclay) at the University of Paris VI Jussieu - Sorbonne in 2001. He was at INSA Lyon holding different positions as a researcher on High-K materials, Centrale Lyon in materials sciences and prospective method and at ESPCI Paris Tech in complex fluids. Dr. LEGRAND joined Varioptic (Parot - Corning) research scientist working on dielectric for electrowetting applications. Dr. LEGRAND joined Radiall in 2010 as manager of Research and Technology corporate leading advanced support on Microwave RF, Thermal mechanical Materials Science and Opto electronic. Dr. LEGRAND is Research & Technology Director since 2019.

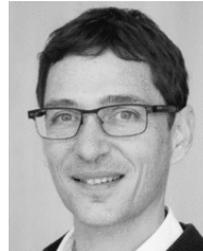

**Dimitri KTENAS** received the Dipl.-Ing. degree in Electrical Engineering from the Ecole Nationale Supérieure d'Electronique et de Radioélectricité (ENSERG), Grenoble (France), in 2001. From that time, he has been with CEA-Leti in Grenoble, France. His main current research interests are PHY, MAC and cross-layer optimization for both B5G cellular networks and LiFi systems. From 2010 to 2015, he was leading the Wireless Communication System Studies laboratory within CEA-Leti, which was in charge of baseband processing and MAC layer studies for wireless systems. Since 2016, he was leading the Broadband Wireless Systems Lab within CEA-Leti, which is in charge of algorithm studies and HW/SW implementation of digital signal processing and protocols for both B5G and LiFi systems. In March 2018, he was appointed Department Head of Wireless Technologies focusing on B5G, IoT and Optical Wireless Communications, from baseband to network layers including propagation modeling, antenna and RF IC design. He has published 60+ scientific papers in international journals and conference proceedings and 5 book chapters, and is the main inventor or co-inventor of 13 patents.